\documentclass[twocolumn]{revtex4}
\usepackage{mathbbold}
\usepackage{amsfonts}
\usepackage{amsmath}
\usepackage{amssymb}
\usepackage{charter}
\usepackage{graphicx}

\input{tcilatex}
\begin{document}

\title{Measurement induced nonclassical states from coherent state heralded
by Knill-Laflamme-Milburn-type SU(3) interference}
\author{Xue--xiang Xu$^{\dag }$ and Shan-jun Ma}
\affiliation{Department of Physics, Jiangxi Normal University, Nanchang 330022, China\\
$^{\dag }$Corresponding author: xuxuexiang@jxnu.edu.cn }

\begin{abstract}
We theoretically generate nonclassical states from coherent state heralded
by Knill-Laflamme-Milburn (KLM)-type SU(3) interference. Injecting a
coherent state in signal mode and two single-photon sources in other two
auxiliary modes of SU(3) interferometry, a broad class of useful
nonclassical states are obtained in the output signal port after making two
single-photon-counting measurements in the two output auxiliary modes. The
nonclassical properties, in terms of anti-bunching effect and squeezing
effect as well as the negativity of the Wigner function, are studied in
detail by adjusting the interaction parameters. The results show that the
input coherent state can be transformed into non-Gaussian states with higher
nonclassicality after measurement induction. The maximum squeezing of our
generated states can be arrived at about 1.9 dB.

\textbf{PACS: }03.67.-a, 05.30.-d, 42.50,Dv, 03.65.Wj

\textbf{Keywords:} nonclassical state; non-Gaussian state; SU(3)
interference; linear optics; mesurement-induced; Wigner function
\end{abstract}

\maketitle

\section{Introduction}

Nonclassical optical states, as the basic sources, play essential roles in
continuous variables (CV) quantum information processing (QIP) \cite{1,2}.
Quantum information over CVs has yielded many exciting advances, both
theoretically and experimently, in fields such as quantum teleportation \cite%
{3,4}, quantum metrology \cite{5}, and potentially quantum computation \cite%
{6,7}. In order to enhance performance to implement various tasks and
provide quantum advantages not attainable classically, a number of
world-wide groups has embarked on route towards understanding, generating,
and ultimately manipulating various nonclassical states of light \cite{8}.
In recent years, many types of nonclassical states have been proposed in
theory and even have already been implemented in laboratories. Agarwal and
Tara generated a nonclassical state by operating photon addition on a
coherent state and investigated the non-classicality of field states \cite{9}%
. Zavatta \textit{et al.} had realized the single-photon added coherent
state \cite{10} and single-photon-added thermal state \cite{11}.

Gaussian states play a prominent roles in CV quantum information \cite{12,13}%
. However, many quantum technological protocols, such as entanglement
distillation \cite{14,15} or quantum error correction \cite{16}, necessarily
require the use of non-Gaussian state, which can extract ultimate potential
of quantum information theory unaccessible by the Gaussian states. With the
increasing significance of non-Gaussian states, the issues of generating
non-Gaussian states have been naturally arisen. It is well-known that the
non-Gaussian states must exhibit non-Gaussian features in their Wigner
distributions in the phase space \cite{17} and gererating non-Gaussian
states normally requires nonlinearities in the field operators \cite{18}. In
order to obtain non-Gaussian states, there is an alternative idea of
measurement-induced scheme (also the conditional probabilistic operation) to
obtain nonlinearity \cite{19}. Some effective nonlinearity \cite{20} is
associated with non-Gaussian measurement such as homodyne measurements or
photon countings \cite{21,22} based on a linear quantum-optical system \cite%
{23}, which generally consist of beam splitting, phase shifting, squeezing,
displacement, and various detection \cite{24}.

Many examples of such measurement-induced schemes by linear optics have been
theoretically implemented, where nonlinear operators are obtained and then
non-Gaussian states are generated \cite{25,26}. Dakna \textit{et al.}
generated a Schrodinger-cat-like state from a single-mode squeezed vacuum
state\ by subtracting photons with low reflectance beam-splitters (BSs) and
photon counters \cite{27}. Subsequently, Lvovsky and Mlynek proposed the
idea of \textquotedblleft quantum-optical catalysis\textquotedblright\ and
generated a coherent superposition state $t\left\vert 0\right\rangle +\alpha
\left\vert 1\right\rangle $ \cite{28}. This state was generated in one of
the BS output channels if a coherent state $\left\vert \alpha \right\rangle $
and a single-photon Fock state $\left\vert 1\right\rangle $ are present in
two input ports and a single photon is registered in the other BS output.
They called this transformation as \textquotedblleft quantum-optical
catalysis\textquotedblright\ because the single photon itself remains
unaffected but facilitates the conversion of the target ensemble. Following
Lvovsky and Mlynek's work and using \textquotedblleft quantum-optical
catalysis\textquotedblright , Bartley \textit{et al.} further generated
multiphoton nonclassical states via interference between coherent and Fock
states, which exhibit a wide range of nonclassical phenomena \cite{29}.
Indeed, the key step in Bartley's scheme is to employ \textquotedblleft
quantum-optical catalysis\textquotedblright\ on the input coherent state one
time. Recently, we operated \textquotedblleft quantum-optical
catalysis\textquotedblright\ on each mode of the two-mode squeezed vacuum
state and generated a non-Gaussian two-mode quantum state with higher
entanglement, accompanied by a nonlinear operator $c_{0}\allowbreak
+c_{1}a^{\dag }b^{\dag }+c_{2}a^{\dag 2}\allowbreak b^{\dag 2}$ \cite{30}.
Enlighten by above works, we further consider the following problem: If we
perform \textquotedblleft quantum-optical catalysis\textquotedblright\ on a
coherent state two times, then what states will be generated and what
nonclassical phenomena will happen? This is the kernal focus of our work.

In 2000, Knill, Laflamme, and Milburn (KLM) implemented a nonlinear sign
(NS) change using a optical network composed of three beam splitters in
succession, whose main features are the use of two ancilla modes with one
prepared photon and post-selection based on measuring the ancillas \cite{31}%
. Because of the SU(3) algebra property of this setup, we name it as
KLM-type SU(3) interferometry. Following KLM's program and using KLM-type
SU(3) interferometry, Ralph \textit{et al.} constructed a conditional
quantum control-NOT gate from linear optical elements \cite{32}. Laterly,
Scheel \textit{et al.} investigated the generation of nonlinear operators
and constructed useful single-mode and two-mode quantum gates necessary for
all-optical QIP \cite{33}. Hence we also use the KLM-type SU(3)
interferometry as our framework to generate nonclassical states.

In this paper, based on the network of KLM-type SU(3) interferometry, we
generate a broad class of nonclassical states from a coherent state by
operating two-fold \textquotedblleft quantum-optical
catalysis\textquotedblright , a kind of measurement-induced scheme. Special
attention is paid to study the nonclassical properties for the generated
states in terms of anti-bunching effect and squeezing effect as well as the
negativity of the Wigner function. We show that the input coherent state can
be transformed into non-Gaussian states with higher nonclassicality after
measurement induction. The paper is organized as follows. In section II, We
briefly outline the KLM-type SU(3) interferometry and introduce the
measurement-induced scheme for generating nonclassical states. In
particular, the normalization factor (i.e. success probability) is
discussed. In section III, we investigate the nonclassical properties
(anti-bunching effect and squeezing effect) of the generated state\ and
analyze the effect of the interaction. Then in section IV we consider the
negativity of the Wigner function, another nonclassical indication for a
quantum state. We summarize our main results in Sec.V.

\section{Measurement induced nonclassical state}

In this section, we briefly outline the dynamics of KLM SU(3)
interferometry\ and apply this interferometry to generate nonclassical
states by measurement-induced protocol. The theoretical scheme is proposed
and the success probability is derived.

\subsection{KLM-type SU(3) interferometry}

Linear optical networks have important applications in QIP and can be
realized in experiment \cite{34}. The quantum mechanical description of
linear optics can be found in Ref \cite{35}. Recently, there has been a
growing interest in linear optical networks in the context of quantum
information technology. It has been demonstrated that universal
(nondeterministic) quantum computation is poissible when linear optical
networks are combined with single photon detectors \cite{36}. Among the
networks, the KLM-type SU(3) interferometry is an important instrument\ in
quantum mechanics and quantum technology.

As illustrated in Fig.1, the main framework of KLM-type SU(3) interferometry
is composed of three beam splitters (BSs) in sequence. The three bosonic
modes containing the photons will be described by creation (annhilation)
operators labeled $a^{\dag }\left( a\right) $, $b^{\dag }\left( b\right) $,
and $c^{\dag }\left( c\right) $. The action of this network can be described
by the unitary operator $U=U_{3}\left( \eta _{3}\right) U_{2}\left( \eta
_{2}\right) U_{1}\left( \eta _{1}\right) $, where $U_{1}\left( \eta
_{1}\right) =e^{\theta _{1}\left( bc^{\dag }-b^{\dag }c\right) }$, $%
U_{2}\left( \eta _{2}\right) =e^{\theta _{2}\left( ba^{\dag }-b^{\dag
}a\right) }$ and $U_{3}\left( \eta _{3}\right) =e^{\theta _{3}\left(
bc^{\dag }-b^{\dag }c\right) }$ are the corresponding operators of the three
BSs with their respective reflection rates $\cos \theta _{j}=\eta _{j}$ ($%
j=1,2,3$). One finds that this effect generates a linear transformation of
the mode operators in the Heisenberg picture. Hence we know the dynamics of
the creation operators:%
\begin{equation}
U\left( 
\begin{array}{c}
a^{\dag } \\ 
b^{\dag } \\ 
c^{\dag }%
\end{array}%
\right) U^{\dag }=\left( 
\begin{array}{ccc}
S_{11} & S_{12} & S_{13} \\ 
S_{21} & S_{22} & S_{23} \\ 
S_{31} & S_{32} & S_{33}%
\end{array}%
\right) \left( 
\begin{array}{c}
a^{\dag } \\ 
b^{\dag } \\ 
c^{\dag }%
\end{array}%
\right) ,  \label{1}
\end{equation}%
Here, the scattering matrix $S$ is a $3\times 3$ one with\ its elements: $%
S_{11}=-\sqrt{\eta _{2}}$, $S_{12}=\sqrt{\eta _{1}\left( 1-\eta _{2}\right) }
$, $S_{13}=\sqrt{\left( 1-\eta _{1}\right) \left( 1-\eta _{2}\right) }$, $%
S_{21}=\sqrt{\left( 1-\eta _{2}\right) \eta _{3}}$, $S_{22}=\sqrt{\left(
1-\eta _{1}\right) \left( 1-\eta _{3}\right) }+\sqrt{\eta _{1}\eta _{2}\eta
_{3}}$, $S_{23}=\sqrt{\left( 1-\eta _{1}\right) \eta _{2}\eta _{3}}-\sqrt{%
\eta _{1}\left( 1-\eta _{3}\right) }$, $S_{31}=\sqrt{\left( 1-\eta
_{2}\right) \left( 1-\eta _{3}\right) }$, $S_{32}=\sqrt{\eta _{1}\eta
_{2}\left( 1-\eta _{3}\right) }-\sqrt{\left( 1-\eta _{1}\right) \eta _{3}}$,
and $S_{33}=\sqrt{\eta _{1}\eta _{3}}+\sqrt{\left( 1-\eta _{1}\right) \eta
_{2}\left( 1-\eta _{3}\right) }$.

\subsection{Scheme\ for generating nonclassical states}

The considered network is actually a three-input and three-output linear
optical system. Three beams of optical field, i.e. a coherent state $%
\left\vert \alpha _{a}\right\rangle $ and two single-photon resources $%
\left\vert 1_{b}\right\rangle $ and $\left\vert 1_{c}\right\rangle $, are
injected into the three-input ports in their respective optical modes $a$, $%
b $, $c$. By the way, we assusme the amplitude of the coherent state $\alpha
=\left\vert \alpha \right\vert e^{i\theta }$ with $\theta =0$ for
simplification. After the interaction in the linear optical system, we make
single-photon-counting measurements in the $b$-mode and $c$-mode output
ports. Thus a conditional quantum state $\left\vert \psi _{a}\right\rangle $
is obtained theoretically and given by%
\begin{equation}
\left\vert \psi _{a}\right\rangle =\frac{1}{\sqrt{p_{d}}}\left\langle
1_{c}\right\vert \left\langle 1_{b}\right\vert U\left\vert \alpha
_{a}\right\rangle \left\vert 1_{b}\right\rangle \left\vert 1_{c}\right\rangle
\label{2}
\end{equation}%
which is generated by measurement induction. The normalization factor $p_{d}$
represents the probability heralded by the successful single-photon
detection in two auxiliary modes. After detailed deviation in appendix A,
the explicit expression of $\left\vert \psi _{a}\right\rangle $\ is given by%
\begin{equation}
\left\vert \psi _{a}\right\rangle =\left( c_{0}+c_{1}a^{\dag }+c_{2}a^{\dag
2}\right) \left\vert S_{11}\alpha \right\rangle ,  \label{3}
\end{equation}%
with the cofficients $c_{0}=\Pi \left( S_{22}S_{33}+S_{23}S_{32}\right) $, $%
c_{1}=\alpha \Pi
(S_{12}S_{21}S_{33}+S_{12}S_{31}S_{23}+S_{21}S_{13}S_{32}+S_{13}S_{22}S_{31}) 
$, $c_{2}=\alpha ^{2}\Pi S_{12}S_{21}S_{13}S_{31}$ and $\Pi =e^{-\left(
1-S_{11}^{2}\right) \left\vert \alpha \right\vert ^{2}/2}/\sqrt{p_{d}}$.
Obviously, a optical operator $c_{0}+c_{1}a^{\dag }+c_{2}a^{\dag 2}$\ is
implemented in this interaction. Note that the coefficients $c_{0}$, $c_{1}$%
, and $c_{2}$ are as the functions of all the relevant parameters. The
interaction parameters involve the coherent-state amplitude $\alpha $, the
beam-splitter reflectance rates $\eta _{1}$, $\eta _{2}$, and $\eta _{3}$.
The generated quantum state can be looked as a non-Gaussian state by
operating this operator on another coherent state $\left\vert S_{11}\alpha
\right\rangle $. Not surprisingly, the input coherent state becomes
non-Gaussian after the process. In particular, the generated state $%
\left\vert \psi _{a}\right\rangle $\ can be reduced to the input coherent
state $\left\vert \alpha _{a}\right\rangle $ when $\eta _{1}=\eta _{2}=\eta
_{3}=\eta \rightarrow 1$. 
\begin{figure}[tbp]
\label{Fig1} \centering\includegraphics[width=1.0\columnwidth]{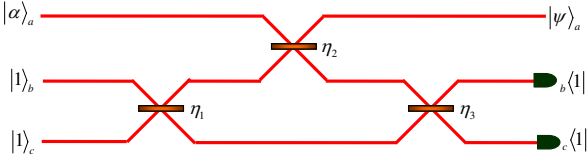}
\caption{(Colour online) Schematic setup of the KLM circuit for generating a
quantum state $\left\vert \protect\psi _{a}\right\rangle $ from an inputting
coherent state $\left\vert \protect\alpha _{a}\right\rangle $ using three
beam splitters (with reflection rates $\protect\eta _{1},\protect\eta _{2},%
\protect\eta _{3}$ respectively), two single-photon sources, and two
single-photon resolving detectors. In contrast with $\left\vert \protect%
\alpha _{a}\right\rangle $, the generated state $\left\vert \protect\psi %
_{a}\right\rangle $ has a wide range of nonclassical properties.}
\end{figure}

By tuning the parameters of the interaction, namely, $\left\vert \alpha
\right\vert $, $\eta _{1}$, $\eta _{2}$, and $\eta _{3}$, the cofficients
may be modulated, generating abroad class of nonclassical states with a wide
range of nonclassical phenomena, as shown next by us.

\subsection{Success probability of detection}

Normalization is important for discussing the properties of a quantum state.
The normalization factor of the generated states in theory is actually the
probability $p_{d}$ of counting successfully single photons at the two
auxiliary modes in experiment. The density operator of the generated state $%
\rho _{a}=\left\vert \psi _{a}\right\rangle \left\langle \psi
_{a}\right\vert $ is expressed in Appendix C. According to $\mathrm{Tr}%
\left( \rho _{a}\right) =1$, the success probability to get the state $%
\left\vert \psi _{a}\right\rangle $\ from our proposal is given by

\begin{eqnarray}
p_{d} &=&(g_{0}+g_{1}\left\vert \alpha \right\vert ^{2}+g_{2}\left\vert
\alpha \right\vert ^{4}  \notag \\
&&+g_{3}\left\vert \alpha \right\vert ^{6}+g_{4}\left\vert \alpha
\right\vert ^{8})e^{-\left( 1-S_{11}^{2}\right) \left\vert \alpha
\right\vert ^{2}},  \label{4}
\end{eqnarray}%
Here $g_{0}$, $g_{1}$, $g_{2}$, $g_{3}$, and $g_{4}$\ are given in appendix
D.

In our following numerical works, the quantities under considering are
discussed only in the two special cases: (1) set $\eta _{1}=\eta _{2}=\eta
_{3}=\eta $ and change $\eta $, (2) set $\eta _{1}=\eta _{3}=1/2$ and change 
$\eta _{2}$. In order to exhibit numerically the probability $p_{d}$, we
plot the contour of $p_{d}$ in the $\left( \eta ,\left\vert \alpha
\right\vert \right) $ plain space in Fig.2 (a) and in the $\left( \eta
_{2},\left\vert \alpha \right\vert \right) $ plain space in Fig.2 (b).
Additionally, we plot $p_{d}$ as a function of $\eta $ in Fig.2 (c) and as a
function of $\eta _{2}$ in Fig.2 (d) for $\left\vert \alpha \right\vert
=1,2,3$. For a large $\left\vert \alpha \right\vert $ and small $\eta $ (or $%
\eta _{2}$), the probability of detection $p_{d}$ is small and even less
than $0.05$. For values closer to $\eta =1$, the generated state gets closer
to the origin input state $\left\vert \alpha \right\rangle $ and the
probability gets closer to $1$. 
\begin{figure}[tbp]
\label{Fig2} \centering\includegraphics[width=1.0\columnwidth]{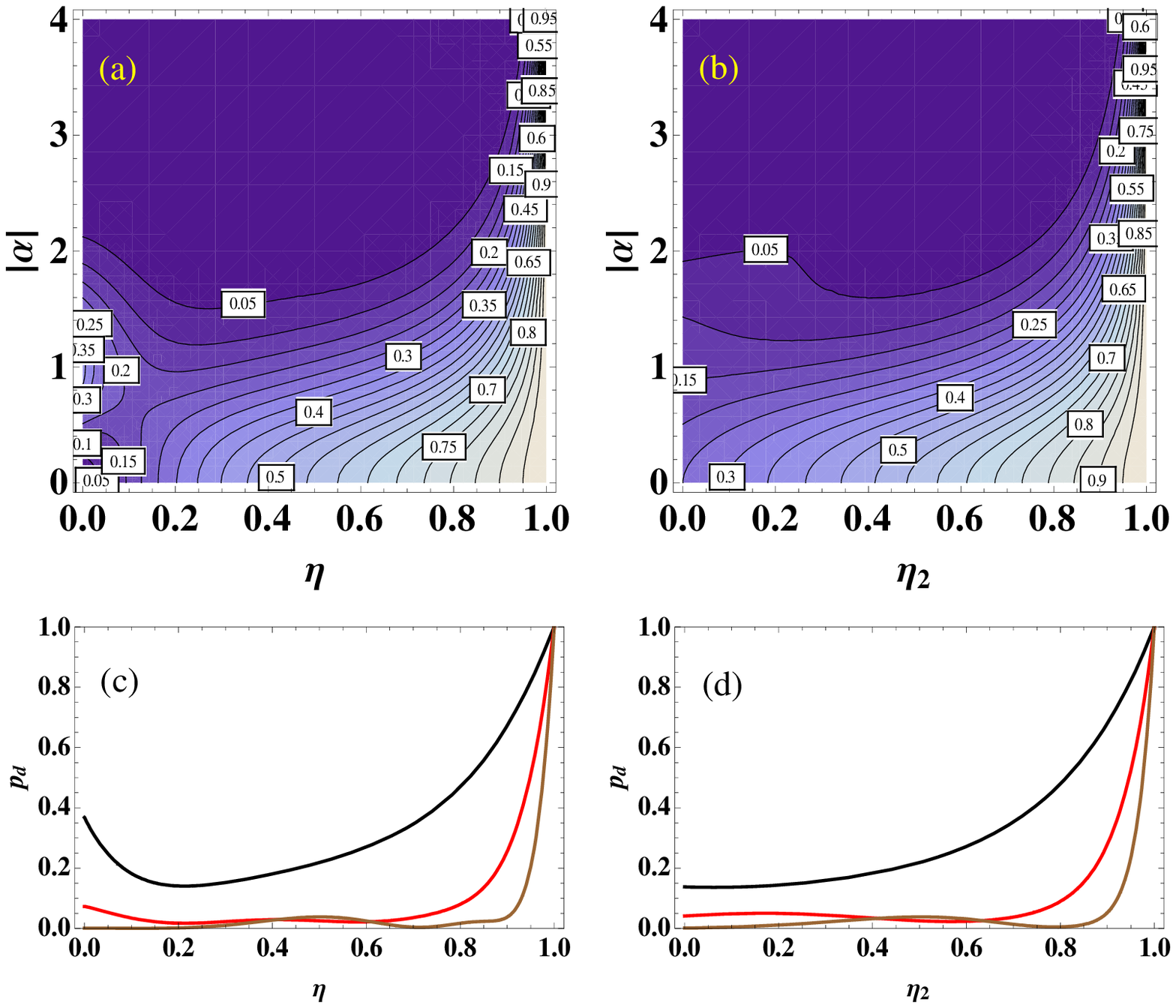}
\caption{(Colour online) Success probability $p_{d}$ of detection (a)
contour plot in ($\protect\eta ,\left\vert \protect\alpha \right\vert $)
space with $\protect\eta =\protect\eta _{1}=\protect\eta _{2}=\protect\eta %
_{3}$; (b) contour plot in ($\protect\eta _{2},\left\vert \protect\alpha %
\right\vert $) space with $\protect\eta _{1}=\protect\eta _{3}=1/2$; (c) as
a fuction of $\protect\eta $; (d) as a fuction of $\protect\eta _{2}$. Here
the black, red and brown lines are corresponding to $\left\vert \protect%
\alpha \right\vert =1$, $\left\vert \protect\alpha \right\vert =2$, $%
\left\vert \protect\alpha \right\vert =3$, respectively. For a large
amplitude of the coherent state $\left\vert \protect\alpha \right\vert $,
the probability is smaller than $0.05$. When $\protect\eta $ is limit to $1$%
, the probability is limit to 1.}
\end{figure}

\section{Nonclassical properties of the generated states}

Quantum states of light can be classified according to their statistical
properties. They are usually compared to a reference state, namely, the
coherent state \cite{37}. In comparison with the input coherent state, what
nonclassical properties will exhibit after meaurement induction in our
proposed scheme. Analytical expressions of the expected values $\left\langle
a^{\dag k}a^{l}\right\rangle $ for arbitrary $k,l$\ found in appendix F
allow us to study\ the statistical properties of this generated states in
our following works. Here, we will focus on studying some nonclassical
properties of this generated quantum state, including anti-bunching effect
and quadrature squeezing effect.

\subsection{Anti-bunching effect}

The second-order autocorrelation function $g^{\left( 2\right) }\left(
0\right) =\left\langle a^{\dagger 2}a^{2}\right\rangle /\left\langle
a^{\dagger }a\right\rangle ^{2}$ determines whether the source produce
effects following antibunching ($g^{\left( 2\right) }\left( 0\right) <1$),
bunching ($1\leqslant g^{\left( 2\right) }\left( 0\right) \leqslant 2$), or
superbunching ($g^{\left( 2\right) }\left( 0\right) >2$). Additionally, it
also determines whether the source produce photons following sub- ($%
g^{\left( 2\right) }\left( 0\right) <1$), super- ($g^{\left( 2\right)
}\left( 0\right) >1$), or Poissonlike ($g^{\left( 2\right) }\left( 0\right)
=1$) statistics \cite{38}. For a coherent state, we have $g^{\left( 2\right)
}\left( 0\right) =1$, which shows its character of Poisson distribution.
Here, we shall examine the anti-bunching effect (a strictly nonclassical
character) of the generated states, which describes whether the photons in
the beam tend to stay apart.

The variations of $g^{\left( 2\right) }\left( 0\right) $ with the
interaction parameters\ are showed in Fig.3. For several given amplitudes of
coherent state ($\left\vert \alpha \right\vert =1,2,3$), we plot $g^{\left(
2\right) }\left( 0\right) $ as a function of $\eta $ in Fig.3 (a)\ and as a
function of $\eta _{2}$ in Fig.3 (b). In a extreme case, we verify that $%
g^{\left( 2\right) }\left( 0\right) =1$ when $\eta $ (or $\eta _{2}$) is
limit to $0$, i.e., the states corresponding to the input coherent state $%
\left\vert \alpha \right\rangle $. Moreover, the feasibility regions for
antibunching, bunching and superbunching are exhibited in the ($\eta
,\left\vert \alpha \right\vert $) parameter space in Fig.3 (c) and in the ($%
\eta _{2},\left\vert \alpha \right\vert $) parameter space in Fig.3 (d). It
is found that there may present antibunching effect in a wide range of
interaction parameters. The results show that the generated state $%
\left\vert \psi _{a}\right\rangle $ can exhibit a broad range of
nonclassical features by tuning the interaction parameter. 
\begin{figure}[tbp]
\label{Fig3} \centering\includegraphics[width=1.0\columnwidth]{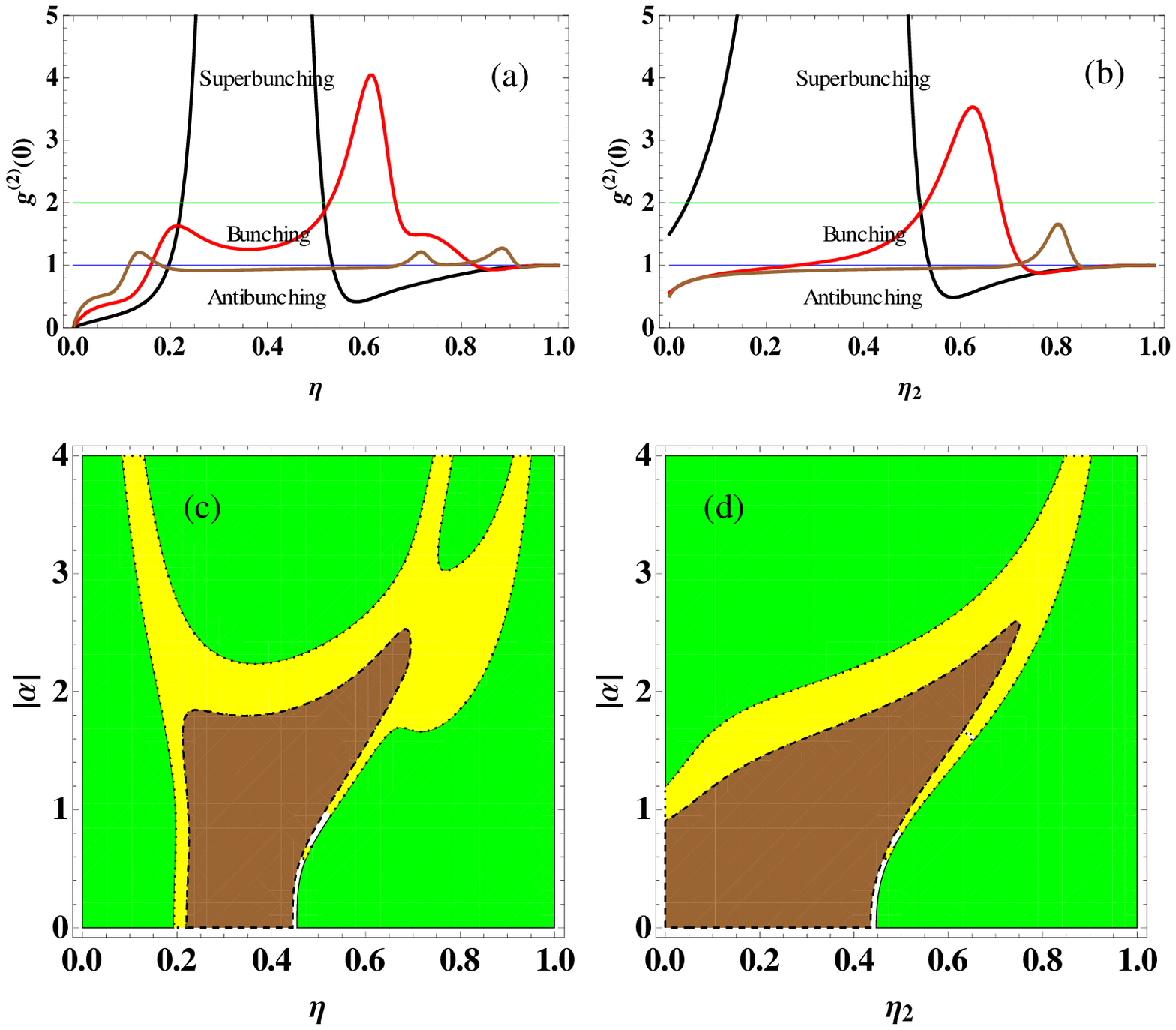}
\caption{(Color online) Second-order autocorrelation function $g^{\left(
2\right) }\left( 0\right) $\ of the generated state. The regions shows
antibunching $g^{\left( 2\right) }\left( 0\right) <1$\ (strictly
nonclassical), bunching $1\leqslant g^{\left( 2\right) }\left( 0\right)
\leqslant 2$, and superbunching $g^{\left( 2\right) }\left( 0\right) >2$.
(a) as a function of $\protect\eta $ ($\protect\eta _{1}=\protect\eta _{2}=%
\protect\eta _{3}=\protect\eta $); (b) as a function of $\protect\eta _{2}$ (%
$\protect\eta _{1}=\protect\eta _{3}=1/2$). Here the black, red and brown
lines are corresponding to $\left\vert \protect\alpha \right\vert =1$, $%
\left\vert \protect\alpha \right\vert =2$, $\left\vert \protect\alpha %
\right\vert =3$, respectively. (c) the feasibility region\ in ($\protect\eta 
$, $\left\vert \protect\alpha \right\vert $) plain space.(d) the feasibility
region\ in ($\protect\eta _{2}$, $\left\vert \protect\alpha \right\vert $)
plain space. Here the blue, yellow and brown regions are corresponding to
antibunching, bunching, and superbunching cases respectively.}
\end{figure}

\subsection{Quadrature squeezing effect}

Squeezed light has come a long way since its first demonstration 30 years 
\cite{39}. Significant advancements have been made from the initial 0.3 dB
squeezing till todys near 13 dB squeezing \cite{40}. Hence we will ask two
questions: (1) Whether our generated states are squeezed states? (2) If they
are squeezed states, then how much can the squeezing degree be arrived? Here
we will consider the squeezing effect of these states.

Firstly we make a brief review of quadrature squeezing effect. Many
experiments have been carried out dealing with noise in the quadrature
component of the field, which is defined by two quadrature operators $%
X=\left( a+a^{\dag }\right) /\sqrt{2}$\ and $P=\left( a-a^{\dag }\right)
/\left( \sqrt{2}i\right) $, analogous to the position and momentum of a
harmonic oscillator \cite{41}. Both quadrature variances, related with the
creation and annhilation operators, can expressed as $\left\langle \Delta
X^{2}\right\rangle =(\left\langle a^{\dag 2}\right\rangle -\left\langle
a^{\dag }\right\rangle ^{2}+\left\langle a^{2}\right\rangle -\left\langle
a\right\rangle ^{2}+1)/2+\left\langle a^{\dag }a\right\rangle -\left\langle
a^{\dag }\right\rangle \left\langle a\right\rangle $ and $\left\langle
\Delta P^{2}\right\rangle =(-\left\langle a^{\dag 2}\right\rangle
+\left\langle a^{\dag }\right\rangle ^{2}-\left\langle a^{2}\right\rangle
+\left\langle a\right\rangle ^{2}+1)/2+\left\langle a^{\dag }a\right\rangle
-\left\langle a^{\dag }\right\rangle \left\langle a\right\rangle $
respectively, as can be seen from their definitions. These components cannot
be measured simulatanously because of the commutation relation $[X,P]=i$. It
follows that the product of the variances in the measurements of the two
quadratures $X$\ and $P$\ satisfies $\Delta X^{2}\Delta P^{2}\geq 1/4$ (a
Heisenberg inequality). For a vacuum state $\left\vert 0\right\rangle $ or a
coherent state $\left\vert \alpha \right\rangle $, the uncertainty relation
is satisfied as an equality, and the two variances are identical: $\Delta
X^{2}|_{\left\vert 0\right\rangle ,\left\vert \alpha \right\rangle }=\Delta
P^{2}|_{\left\vert 0\right\rangle ,\left\vert \alpha \right\rangle }=1/2$.\
A quantum state is called squeezed if the variance of a quadrature amplitude
is below the variance of a vacuum or a coherent state, i.e. $\Delta
X^{2}<1/2 $ or $\Delta P^{2}<1/2$. The squeezing effect of a light field
comes at the expense of increasing the fluctuations in the other quadrature
amplitude. Here we can adopt quantum squeezing quantified in a dB scale
through $dB[X]=10\log _{10}\left( \Delta X^{2}/\Delta X^{2}|_{\left\vert
0\right\rangle }\right) $, $dB[P]=10\log _{10}\left( \Delta P^{2}/\Delta
P^{2}|_{\left\vert 0\right\rangle }\right) $. In other words, if $dB[X]$\ or 
$dB[P]$\ is negative, this quantum state is a squeezed state.

In Fig.4 (a), the behaviour of $dB[X]$ as a function of $\eta $ ($=\eta
_{1}=\eta _{2}=\eta _{3}$) for different $\left\vert \alpha \right\vert $.
By minimizing the expression of $\Delta X^{2}$, the largest squeezing
attained is around $0.321772$, below the vacuum noise level of $1/2$ by
approximately $1.91422$ dB\ (using MATHENATICA) for $\left\vert \alpha
\right\vert =1$. The squeezed degree is bigger than that ($1.25$ dB) in Ref.%
\cite{29}. In Fig.4 (b), the behaviour of $dB[X]$ as a function of $\eta
_{2} $ for different $\left\vert \alpha \right\vert $ with $\eta _{1}=\eta
_{3}=1/2$. Moreover, the purple regions show the feasibility squeezed region
of $X$ quadrature component in the ($\eta ,\left\vert \alpha \right\vert $)
parameter space in Fig.4 (c) and in the ($\eta _{2},\left\vert \alpha
\right\vert $) parameter space in Fig.4 (d). Fig.4 (c) and 4 (d) show a wide
range of squeezing for $\left\vert \alpha \right\vert $ and $\eta \left(
\eta _{2}\right) $.

\begin{figure}[tbp]
\label{Fig4} \centering\includegraphics[width=1.0\columnwidth]{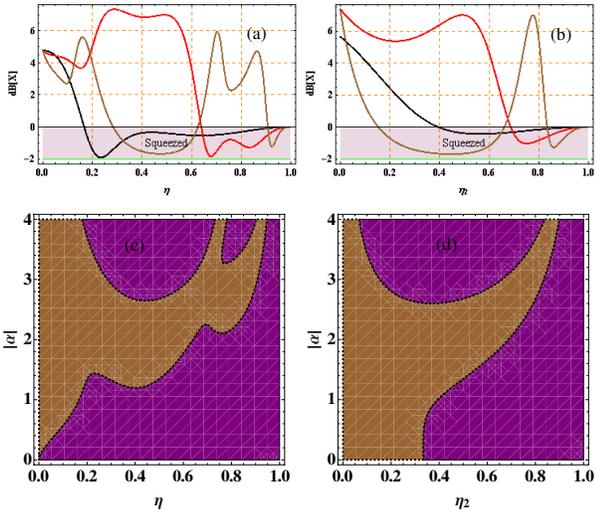}
\caption{(Color online) Quadrature variance of $X$ component relative to the
vacuum (unsqueezed) state in units of dB (a) as a function of $\protect\eta $
($\protect\eta _{1}=\protect\eta _{2}=\protect\eta _{3}=\protect\eta $); (b)
as a function of $\protect\eta _{2}$ ($\protect\eta _{1}=\protect\eta %
_{3}=1/2$). Here the black, red and brown solid lines are corresponding to $%
\left\vert \protect\alpha \right\vert =1$, $\left\vert \protect\alpha %
\right\vert =2$, $\left\vert \protect\alpha \right\vert =3$ in X component,
respectively. The feasibility squeezed (purple) and unsqueezed (brown)
region of the $X$ component (c)\ in ($\protect\eta $, $\left\vert \protect%
\alpha \right\vert $) plain space; (d) in ($\protect\eta _{2}$, $\left\vert 
\protect\alpha \right\vert $) plain space.}
\end{figure}

\section{Wigner function of the generated states}

The negative Wigner function is a witness of the nonclassicality of a
quantum state \cite{42}. Additionaly, we can determinate whether this
quantum state is non-Gaussian state from the form of the Wigner function 
\cite{43}. Since non-Gaussian quantum state can provide quantum advantages
not attainable classically, it is necessary to study the Wigner functions of
our generated states.

The analytical expression of the Wigner function for the generated states is
derived in appendix F. The results plotted in Fig.5 are obtained for optimal
choices with different parameters $\left( \left\vert \alpha \right\vert
,\eta _{1},\eta _{2},\eta _{3}\right) $. Fig.5 shows that the Wigner
functions are negative in some regions of the phase space, which\ is a
witness of the nonclassicality. As we all know, the coherent state is a
typical Gaussian state whose Wigner function has no negative regions. In
comparison with the input coherent state, the generated states show the
non-Gaussian features with negative regions in moderate parameters range.

Meanwhile, we plot the negative volume $\delta $\ as a function of $\eta $\
(or $\eta _{2}$) for $\left\vert \alpha \right\vert =1$, $2$, $3$ in Fig.6.
It is obvious to see that for different $\left\vert \alpha \right\vert $,
the maximun negative volumn locate at different $\eta $\ (or $\eta _{2}$).
For instance, when $\left\vert \alpha \right\vert =1$, $\delta _{max}$ is
located at $\eta =0$\ (or $\eta _{2}=0$) and $\delta $ is decreaing with $%
\eta $\ (or $\eta _{2}$) increasing and limit to zero at large $\eta $\ (or $%
\eta _{2}$); For $\left\vert \alpha \right\vert =2$, $\delta _{max}$ is
located at around $\eta =0.27$\ (or $\eta _{2}=0.5$); For $\left\vert \alpha
\right\vert =3$, $\delta _{max}$ is located at around $\eta =0.7$\ (or $\eta
_{2}=0.8$). 
\begin{figure}[tbp]
\label{Fig5} \centering\includegraphics[width=1.0\columnwidth]{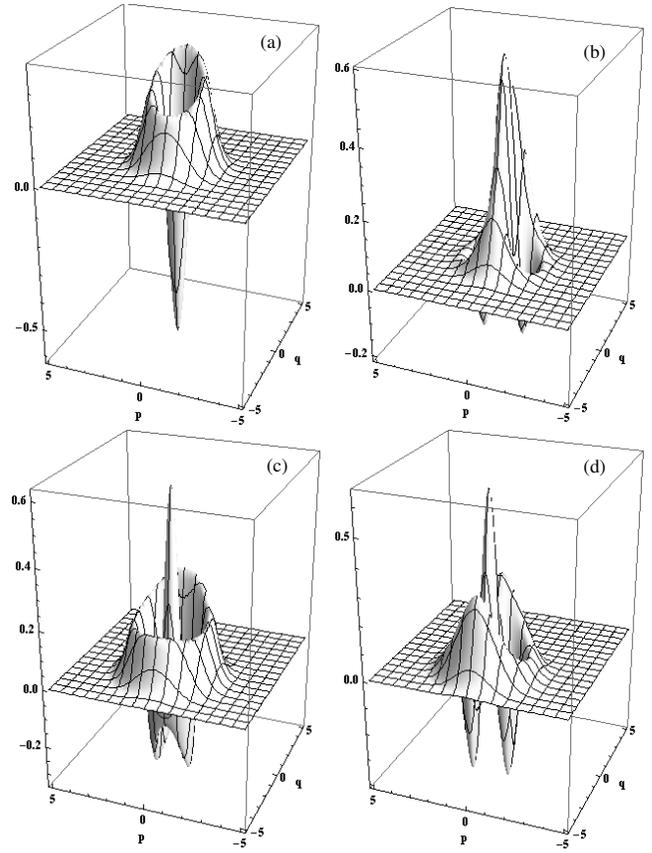}
\caption{(Color online) Wigner functions of the measurement-induced state
with parameters $\left( \left\vert \protect\alpha \right\vert ,\protect\eta %
_{1},\protect\eta _{2},\protect\eta _{3}\right) $: (a) $\left(
1,0.05,0.05,0.05\right) $; (b) $\left( 2,0.5,0.05,0.5\right) $; (c) $\left(
2,0.25,0.25,0.25\right) $; (d) $\left( 2,0.5,0.5,0.5\right) $. The result
shows that the Wigner function of the generated state has the negative
regions, which is different from that of the input coherent state and also
shows the nonclassical character.}
\end{figure}
\begin{figure}[tbp]
\label{Fig6} \centering\includegraphics[width=1.0\columnwidth]{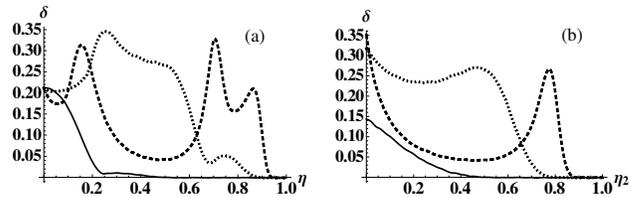}
\caption{(Color online) The negative volumn $\protect\delta $ of the Wigner
functions of the generated states (a) as a function of $\protect\eta $ ($%
\protect\eta _{1}=\protect\eta _{2}=\protect\eta _{3}=\protect\eta $); (b)
as a function of $\protect\eta _{2}$ ($\protect\eta _{1}=\protect\eta %
_{3}=1/2$). The solid, dotted and dashed line are corresponding to $%
\left\vert \protect\alpha \right\vert =1$, $\left\vert \protect\alpha %
\right\vert =2$, $\left\vert \protect\alpha \right\vert =3$, respectively.}
\end{figure}

\section{Discussion and Conclusion}

Our results show that there exist optimal nonclassical properites in
different parameter ranges. The optimal performance as required can be
obtained by adjusting the interaction parameters. A prominent character is
that the the maxinum squeezing can be reached to $1.91422$ dB. In comparison
with the states generated in Bartley's work, the squeezing degree of our
generated states is enhanced. In table I, we numerate some values of the
probability of detection $p_{d}$, the auto-correlation function $g^{\left(
2\right) }\left( 0\right) $, the squeezing in $X$ component $dB[X]$, and the
negative volume $\delta $ of the Wigner functions in their corresponding $%
\left( \left\vert \alpha \right\vert ,\eta _{1},\eta _{2},\eta _{3}\right) $
parameters.

\begin{table}[h]
\caption{Values of the probability of detection $p_{d}$, the
auto-correlation function $g^{\left( 2\right) }\left( 0\right) $, the
squeezing in $X$ component $dB[X]$, and the negative volume $\protect\delta $
of the WFs in their corresponding interaction parameter values of $\left(
\left\vert \protect\alpha \right\vert ,\protect\eta _{1},\protect\eta _{2},%
\protect\eta _{3}\right) $. }
\label{result_tableau}
\begin{center}
\begin{tabular}{|c|c|c|c|c|}
\hline\hline
$\left( \left\vert \alpha \right\vert ,\eta _{1},\eta _{2},\eta _{3}\right) $
& $p_{d}$ & $g^{\left( 2\right) }\left( 0\right) $ & $dB\left[ X\right] $ & $%
\delta $ \\ \hline\hline
$\left( 1,0.3,0.3,0.3\right) $ & 0.15223 & 26.0347 & -1.16685 & 0.0103 \\ 
$\left( 1,0.6,0.6,0.6\right) $ & 0.27011 & 0.43187 & -0.52375 & 0.0010 \\ 
$\left( 1,0.9,0.9,0.9\right) $ & \hspace{1pt}0.67275\hspace{1pt} & 0.97122 
\hspace{1pt} & -0.08971 & 0.0009 \hspace{1pt} \\ 
$\left( 1,0.5,0.7,0.5\right) $ & 0.35303 & 0.76685 & -0.31669 & 0.0009 \\ 
$\left( 2,0.5,0.7,0.5\right) $ & 0.03770 & 1.47059 & -0.43680 & 0.0334 \\ 
$\left( 3,0.5,0.7,0.5\right) $ & 0.01392 & 0.99035 & 1.80934 & 0.1283 \\ 
\hline
\end{tabular}%
\end{center}
\end{table}

In summary, this paper employs two-fold quantum-optical catalysis to
generate a class of nonclassical states based on KLM-type SU(3)
interferometry. By measurement induction, we implemente a nonlinear operator 
$c_{0}+c_{1}a^{\dag }+c_{2}a^{\dag 2}$\ and obtain a broad class of useful
non-Gaussian quantum states with higher nonclassicality. We discuss the
success probability of detection and the nonclassical properties in terms of
anti-bunching effect and squeezing effect as well as the negativity of the
Wigner function. In compared with the input coherent state, these generated
nonclassical states exhibit a lot of nonclassical properties. The results
show that the optimal antibunching and the maximum squeezing as well as the
maximum negative volume of the Wigner functions are located in different
parameter points. Hence one can choose the optimal performance of these
nonclassical properties to implement various technological tasks. In
addition, the generation of these nonclassical states is feasible with
current technology. The experiment realization of such states is desired to
achieve in the future. Our results can provid a theoretical reference for
experiments.

\begin{acknowledgments}
This work was supported by the National Nature Science Foundation of China
(Grants No. 11264018 and No. 11447002) and the Natural Science Foundation of
Jiangxi Province of China (Grants No. 20142BAB202001 and No. 20151BAB202013)
\end{acknowledgments}

\textbf{Appendix A: Transformation relation of the SU(3) interferometry}

The network can be seen as three multiports in cascade, one for each of the
beam splitters with the unitary operators $U_{j}\left( \eta _{j}\right) $
and the scattering matrices $S_{j}$ $\left( j=1,2,3\right) $. For each
stage, the dyanamics is $U_{j}\left( a^{\dag },b^{\dag },c^{\dag }\right)
^{T}U_{j}^{\dag }=S_{j}\left( a^{\dag },b^{\dag },c^{\dag }\right) ^{T}$,
with the scattering metrix 
\begin{equation*}
S_{1}=\left( 
\begin{array}{ccc}
1 & 0 & 0 \\ 
0 & \sqrt{\eta _{1}} & \sqrt{1-\eta _{1}} \\ 
0 & \sqrt{1-\eta _{1}} & -\sqrt{\eta _{1}}%
\end{array}%
\right) ,
\end{equation*}%
\begin{equation*}
S_{2}=\left( 
\begin{array}{ccc}
-\sqrt{\eta _{2}} & \sqrt{1-\eta _{2}} & 0 \\ 
\sqrt{1-\eta _{2}} & \sqrt{\eta _{2}} & 0 \\ 
0 & 0 & 1%
\end{array}%
\right) ,
\end{equation*}%
\begin{equation*}
S_{3}=\left( 
\begin{array}{ccc}
1 & 0 & 0 \\ 
0 & \sqrt{\eta _{3}} & \sqrt{1-\eta _{3}} \\ 
0 & \sqrt{1-\eta _{3}} & -\sqrt{\eta _{3}}%
\end{array}%
\right) ,
\end{equation*}%
\ So the total operator is $U=U_{3}\left( \eta _{3}\right) U_{2}\left( \eta
_{2}\right) U_{1}\left( \eta _{1}\right) $ and the total scattering matrix
is $S=S_{3}S_{2}S_{1}$. Thus Eq.(\ref{1}) is obtained. One can also refer
the detailed information in Ref.\cite{35}.

\textbf{Appendix B: Explicit expression of the generated state}

Substituting $\left\vert \alpha _{a}\right\rangle =e^{-\left\vert \alpha
\right\vert ^{2}/2+\alpha a^{\dag }}\left\vert 0_{a}\right\rangle $, $%
\left\vert 1_{b}\right\rangle =\frac{d}{ds_{1}}e^{s_{1}b^{\dag }}\left\vert
0_{b}\right\rangle |_{s_{1}=0}$, $\left\vert 1_{c}\right\rangle =\frac{d}{%
ds_{2}}e^{s_{2}c^{\dag }}\left\vert 0_{c}\right\rangle |_{s_{2}=0}$, $%
\left\langle 1_{b}\right\vert =\frac{d}{ds_{3}}\left\langle 0_{b}\right\vert
e^{s_{3}b}|_{s_{3}=0}$ and $\left\langle 1_{c}\right\vert =\frac{d}{ds_{4}}%
\left\langle 0_{c}\right\vert e^{s_{4}c}|_{s_{4}=0}$, into Eq.(\ref{2}) and
using the transformation in Eq.(\ref{1}) as well as the fact $U\left\vert
0_{a}\right\rangle \left\vert 0_{b}\right\rangle \left\vert
0_{c}\right\rangle =\left\vert 0_{a}\right\rangle \left\vert
0_{b}\right\rangle \left\vert 0_{c}\right\rangle $, we finally arrive at the
derivative form of $\left\vert \psi _{a}\right\rangle $,%
\begin{eqnarray*}
\left\vert \psi _{a}\right\rangle &=&\frac{e^{-\left\vert \alpha \right\vert
^{2}/2}}{\sqrt{p_{d}}}\frac{d^{4}}{ds_{4}ds_{3}ds_{2}ds_{1}}e^{\alpha \left(
S_{12}s_{3}+S_{13}s_{4}\right) } \\
&&\times e^{S_{22}s_{1}s_{3}+S_{23}s_{1}s_{4}+S_{32}s_{2}\allowbreak
s_{3}+S_{33}s_{2}s_{4}} \\
&&\times e^{a^{\dag }\left( \alpha S_{11}+S_{21}s_{1}+S_{31}s_{2}\right)
}\left\vert 0_{a}\right\rangle |_{s_{1}=s_{2}=s_{3}=s_{4}=0},
\end{eqnarray*}%
Therefore the explicit form in Eq.(\ref{2}) can be obtained after making
derivation.

\textbf{Appendix C: Density operator of the generated state}

The conjugate state of $\left\vert \psi _{a}\right\rangle $\ can be given by

\begin{eqnarray*}
\left\langle \psi _{a}\right\vert &=&\frac{e^{-\left\vert \alpha \right\vert
^{2}/2}}{\sqrt{p_{d}}}\frac{d^{4}}{dh_{4}dh_{3}dh_{2}dh_{1}}e^{\alpha ^{\ast
}\left( S_{12}h_{3}+S_{13}h_{4}\right) } \\
&&\times e^{S_{22}h_{1}h_{3}+S_{23}h_{1}h_{4}+S_{32}h_{2}\allowbreak
h_{3}+S_{33}h_{2}h_{4}} \\
&&\times \left\langle 0_{a}\right\vert e^{a\left( \alpha ^{\ast
}S_{11}+S_{21}h_{1}+S_{31}h_{2}\right) }|_{h_{1}=h_{2}=h_{3}=h_{4}=0},
\end{eqnarray*}%
Then, the density operator is%
\begin{eqnarray*}
\rho _{a} &=&\frac{e^{-\left\vert \alpha \right\vert ^{2}}}{p_{d}}\frac{d^{8}%
}{dh_{4}dh_{3}dh_{2}dh_{1}ds_{4}ds_{3}ds_{2}ds_{1}} \\
&&\times e^{\alpha \left( S_{12}s_{3}+S_{13}s_{4}\right) +\alpha ^{\ast
}\left( S_{12}h_{3}+S_{13}h_{4}\right) } \\
&&\times e^{S_{22}s_{1}s_{3}+S_{23}s_{1}s_{4}+S_{32}s_{2}\allowbreak
s_{3}+S_{33}s_{2}s_{4}} \\
&&\times e^{S_{22}h_{1}h_{3}+S_{23}h_{1}h_{4}+S_{32}h_{2}\allowbreak
h_{3}+S_{33}h_{2}h_{4}} \\
&&\times e^{a^{\dag }\left( \alpha S_{11}+S_{21}s_{1}+S_{31}s_{2}\right)
}\left\vert 0_{a}\right\rangle \left\langle 0_{a}\right\vert e^{a\left(
\alpha ^{\ast }S_{11}+S_{21}h_{1}+S_{31}h_{2}\right) } \\
&&|_{s_{1}=s_{2}=s_{3}=s_{4}=h_{1}=h_{2}=h_{3}=h_{4}=0}
\end{eqnarray*}

\textbf{Appendix D: Success probability }$p_{d}$\textbf{\ of detection}

Due to $\mathrm{tr}\left( \rho _{a}\right) =1$, we have%
\begin{eqnarray*}
p_{d} &=&e^{-\left( 1-S_{11}^{2}\right) \left\vert \alpha \right\vert ^{2}}%
\frac{d^{8}}{dh_{4}dh_{3}dh_{2}dh_{1}ds_{4}ds_{3}ds_{2}ds_{1}} \\
&&\times e^{\left( S_{11}S_{21}\allowbreak
h_{1}+S_{11}S_{31}h_{2}+S_{12}s_{3}+S_{13}s_{4}\right) \alpha } \\
&&\times e^{\left(
S_{11}S_{21}s_{1}+S_{11}S_{31}s_{2}+S_{12}h_{3}+S_{13}h_{4}\right) \alpha
^{\ast }} \\
&&\times e^{S_{22}\left( s_{1}s_{3}+h_{1}h_{3}\right) +S_{23}\left(
s_{1}s_{4}+h_{1}h_{4}\right) +S_{21}^{2}s_{1}h_{1}+S_{31}^{2}s_{2}h_{2}} \\
&&\times e^{S_{32}\left( s_{2}\allowbreak s_{3}+h_{2}\allowbreak
h_{3}\right) +S_{33}\left( s_{2}s_{4}+h_{2}h_{4}\right) +S_{21}S_{31}\left(
s_{2}h_{1}+\allowbreak s_{1}h_{2}\right) } \\
&&|_{s_{1}=s_{2}=s_{3}=s_{4}=h_{1}=h_{2}=h_{3}=h_{4}=0}
\end{eqnarray*}%
which lead to the result in Eq.(\ref{4}) with%
\begin{equation*}
g_{0}=\left( S_{23}S_{32}+S_{22}S_{33}\right) ^{2},
\end{equation*}%
\begin{eqnarray*}
g_{1}
&=&(S_{13}S_{22}S_{31}+S_{12}S_{23}S_{31}+S_{13}S_{21}S_{32}+S_{12}S_{21}S_{33})
\\
&&\times \left(
S_{13}S_{22}S_{31}+S_{12}S_{23}S_{31}+S_{13}S_{21}S_{32}\right. \\
&&\left. +2S_{11}S_{23}S_{32}+S_{12}S_{21}S_{33}+2S_{11}S_{22}S_{33}\right) ,
\end{eqnarray*}%
\begin{eqnarray*}
g_{2} &=&S_{11}^{2}S_{13}^{2}\left( S_{22}S_{31}+S_{21}S_{32}\right) ^{2} \\
&&+4S_{11}S_{12}^{2}S_{12}^{2}S_{13}S_{21}S_{31}\left(
S_{23}S_{31}+S_{21}S_{33}\right) \\
&&+2S_{12}^{2}S_{13}^{2}S_{21}^{2}S_{31}^{2}+S_{11}^{2}S_{12}^{2}\left(
S_{23}S_{31}+S_{21}S_{33}\right) ^{2} \\
&&+4S_{11}S_{12}S_{13}S_{13}S_{21}S_{31}\left(
S_{22}S_{31}+S_{21}S_{32}\right) \\
&&+2S_{11}^{2}S_{12}S_{13}S_{21}S_{32}\left(
2S_{23}S_{31}+S_{21}S_{33}\right) \\
&&+2S_{11}^{2}S_{12}S_{13}S_{22}S_{31}\left(
S_{23}S_{31}+2S_{21}S_{33}\right) ,
\end{eqnarray*}%
\begin{eqnarray*}
g_{3} &=&2S_{11}^{3}S_{12}S_{13}^{2}S_{21}S_{31}\left(
S_{22}S_{31}+S_{21}S_{32}\right) \\
&&+4S_{11}^{2}S_{12}^{2}S_{13}^{2}S_{21}^{2}S_{31}^{2}+2S_{11}^{3}S_{12}^{2}S_{13}^{2}S_{21}S_{31}S_{23}
\\
&&+2S_{11}^{3}S_{12}^{2}S_{13}S_{21}^{2}S_{31}S_{33},
\end{eqnarray*}%
\begin{equation*}
g_{4}=S_{11}^{4}S_{12}^{2}S_{13}^{2}S_{21}^{2}S_{31}^{2}.
\end{equation*}

\textbf{Appendix E: General expressions of the expected values }$%
\left\langle a^{\dag k}a^{l}\right\rangle $

According to $\left\langle a^{\dag k}a^{l}\right\rangle =\mathrm{tr}\left(
a^{\dag k}a^{l}\rho _{a}\right) $ and making detailed calculation, we obtain%
\begin{eqnarray*}
&&\left\langle a^{\dag k}a^{l}\right\rangle \\
&=&\frac{e^{-\left( 1-S_{11}^{2}\right) \left\vert \alpha \right\vert ^{2}}}{%
p_{d}}\frac{d^{8+k+l}}{dh_{4}dh_{3}dh_{2}dh_{1}ds_{4}ds_{3}ds_{2}ds_{1}d\mu
^{k}d\nu ^{l}} \\
&&\times e^{\left(
S_{11}S_{21}h_{1}+S_{11}S_{31}h_{2}+S_{12}s_{3}+S_{13}s_{4}\right) \alpha }
\\
&&\times e^{\left( S_{11}S_{21}s_{1}+\allowbreak
S_{11}S_{31}s_{2}+S_{12}h_{3}+S_{13}h_{4}\right) \alpha ^{\ast }} \\
&&\times e^{+S_{22}\left( s_{1}s_{3}+h_{1}h_{3}\right) +S_{23}\left(
s_{1}s_{4}+h_{1}h_{4}\right) +S_{21}^{2}s_{1}h_{1}+S_{31}^{2}s_{2}h_{2}} \\
&&\times e^{+S_{32}\left( s_{2}\allowbreak s_{3}+h_{2}\allowbreak
h_{3}\right) +S_{33}\left( s_{2}s_{4}+h_{2}h_{4}\right) +S_{21}\allowbreak
S_{31}\left( s_{2}h_{1}+s_{1}h_{2}\right) } \\
&&\times e^{S_{11}\left( \mu \alpha ^{\ast }+\nu \alpha \right) +\mu \left(
S_{21}h_{1}+S_{31}h_{2}\right) +\nu \left( S_{21}\allowbreak
s_{1}+S_{31}s_{2}\right) } \\
&&|_{s_{1}=s_{2}=s_{3}=s_{4}=h_{1}=h_{2}=h_{3}=h_{4}=\mu =\nu =0}.
\end{eqnarray*}%
Using this general expression, we can study the statistical properties of
our generated states.

\textbf{Appendix F: Wigner function of the generated state}

According to the formula of the Wigner function in the coherent state
representation $\left\vert z\right\rangle $, i.e $W(\beta )=\frac{%
2e^{2\left\vert \beta \right\vert ^{2}}}{\pi }\int \frac{d^{2}z}{\pi }%
\left\langle -z\right\vert \rho \left\vert z\right\rangle e^{-2\left( z\beta
^{\ast }-z^{\ast }\beta \right) }$ with $\beta =\left( q+ip\right) /\sqrt{2}$%
, the Wigner function is given by 
\begin{eqnarray*}
&&W(\beta ) \\
&=&\frac{2e^{-\left( 1-S_{11}^{2}\right) \left\vert \alpha \right\vert
^{2}-2\left\vert \beta -S_{11}\alpha \right\vert ^{2}}}{\pi p_{d}}\frac{d^{8}%
}{dh_{4}dh_{3}dh_{2}dh_{1}ds_{4}ds_{3}ds_{2}ds_{1}} \\
&&\times e^{\alpha \left(
-h_{1}S_{11}S_{21}-h_{2}S_{11}S_{31}+S_{12}s_{3}+S_{13}s_{4}\right) } \\
&&\times e^{\alpha ^{\ast }\left( -S_{11}S_{21}s_{1}\allowbreak
-S_{11}S_{31}s_{2}+S_{12}h_{3}+S_{13}h_{4}\right) } \\
&&\times e^{+S_{22}\left( s_{1}s_{3}+h_{1}h_{3}\right) +S_{23}\left(
s_{1}s_{4}+h_{1}h_{4}\right) -S_{21}^{2}s_{1}h_{1}-S_{31}^{2}s_{2}h_{2}} \\
&&\times e^{+S_{32}\left( s_{2}\allowbreak s_{3}+h_{2}\allowbreak
h_{3}\right) +S_{33}\left( s_{2}s_{4}+h_{2}h_{4}\right) -S_{21}S_{31}\left(
s_{2}h_{1}+s_{1}h_{2}\right) } \\
&&\times e^{2\beta \left( h_{1}S_{21}+h_{2}S_{31}\right) +2\beta ^{\ast
}\left( S_{21}s_{1}+S_{31}\allowbreak s_{2}\right) } \\
&&|_{s_{1}=s_{2}=s_{3}=s_{4}=h_{1}=h_{2}=h_{3}=h_{4}=0}.
\end{eqnarray*}%
After derivative, the analytical expression can be obtained.

\end{document}